\documentclass{article}
\usepackage{epsfig}
\usepackage[centertags]{amsmath}
\usepackage{graphicx}
\usepackage{url}

\begin{document}
\title{Quantum Control  of Interacting Bosons in Periodic Optical Lattice}
\author{Analabha Roy and L.E. Reichl \\
Center for Complex Quantum Systems\\
and\\
Department of Physics\\
The University of Texas at Austin, Austin, Texas 78712\\}
\date{\today }
\maketitle

\begin{abstract}

We study the avoided crossings in the dynamics of quantum controlled excitations for an interacting two-boson system in an optical lattice.  Specifically, we perform numerical simulations of quantum control in this system where  driving pulses connect the undriven stationary states in a manner characteristic of Stimulated Raman Adiabatic Passage (STIRAP). We demonstrate that the dynamics of such a transition is affected by chaos induced avoided crossings, resulting in a loss in coherence  of the final outcome in the adiabatic limit.

\end{abstract}
\section{Introduction}
\label{sec:intro}

Stimulated Raman Adiabatic Passage, or STIRAP,  is a well-known method of inducing coherent excitations of quantum systems from the ground state to states with higher energy. This is achieved using coherent time-modulated laser fields that result in complete population transfer from an initially populated ground state to a target state via an intermediate state.  STIRAP was first proposed by Hioe and coworkers \cite{stirap:hioe}, \cite{hioe:rwa:stirap}.  A crucial preliminary work by Becker  et al.~\cite{stirap:seminal} achieved efficient vibrational excitation by using a molecular beam as an optically pumped active medium and led to the development of the STIRAP concept. The theoretical work was formulated by Kuklinsky, Gaubatz, Hioe and Bergmann shortly thereafter \cite{stirap:theory}. STIRAP was further confirmed by manipulating the vibrational and rotational degrees of freedom in sodium dimers \cite{stirap:experiment}. 

Since then, STIRAP has been used to control transitions in a diverse range of matter-optics systems \cite{stirap:review}, ranging from molecular alignment \cite{stirap:molecular}, and molecular rotation \cite{na-reichl:mol-rot}, to the coherent acceleration of ultracold atom systems by transfer between a stationary and a moving optical lattice \cite{holder:reichl:2res}, and controlled dipole excitations in ultracold bosons subjected to radiation induced double well potentials  \cite{roy:reichl:dblwell}. In the latter  two cases, STIRAP has also been used to illustrate the influence of the underlying classical chaos in the atom dynamics by looking at avoided crossings in the Floquet eigenphase spectrum. The exact procedure was first described by Na and Reichl  \cite{na-reichl:pbox} for a driven single particle in a box.

Optical lattice systems have been of great interest in experimental physics \cite{oplattice:oldexpt} and \cite{oplattice:oldexpt2}) and theoretical physics \cite{oplattice:oldtheory}, \cite{oplattice:oldtheory2} for some time. In $1992$, Graham, Schlautmann and Zoller showed that the center of mass motion of cold atoms in an optical lattice could be decoupled from their internal degrees of freedom  \cite{graham}.  Since then, the influence of chaos in optical lattice systems has been  an important factor in the manipulation of such systems   \cite{holder:reichl:2res}, \cite{raizen:oplattice}. Number squeezing and subpoissonian distribution of atoms in each site in an optical lattice have been reported by Itah et al.  \cite{technion:oplattice-culling}. The system of interest in our work is an optical lattice with a 2-periodic boundary condition that traps two interacting bosons. We seek to use this system as a test case for larger many-particle problems, where the Hilbert space can be restricted to the subspace of all $N$-periodic wavefunctions, with large $N$. Here, we set $N$ to the lowest meaningful value of $2$, since singly periodic wavefunctions do not take into account the probability of two particles being in separate wells of the lattice. Furthermore, Fernholtz et al.  \cite{fernholtz} have recently  shown that it is possible to trap a cold atomic system on the surface of a toroid and to achieve a two-dimensional periodic potential similar to the ringed optical lattice that we shall consider in this work. 

In subsequent sections, we consider the application of STIRAP to an interacting boson system confined to a one-dimensional optical lattice with periodic boundary conditions.   A linear version of our system has been implemented in  experimental studies~\cite{steck} using ultra-cold atoms. In section~\ref{sec:basicmodel}, we  derive an expression for the basic model and we  discuss the numerical methods used to obtain the stationary eigenstates for this system.  In section~\ref{sec:qmechint}, we discuss the process by which coherent transitions between selected symmetrized energy eigenstates can be achieved for this system.  We will also show that avoided crossings in the Floquet states can be associated with real transitions of the undriven symmetrized eigenstates. Concluding remarks are made in section~\ref{conclusion}.

\section{The Basic Model}
\label{sec:basicmodel}

Our system consists of two  atoms (bosons), each  of mass $m$, confined to a spatially periodic  optical lattice of radius $\rho$.  The dipole interaction between the atom and the optical lattice gives us the  atomic Hamiltonian
\begin{equation}
H=\frac{L_1^2}{2I}+\frac{L_2^2}{2I}+\kappa_0 [ \cos{(2{\theta}_1)} + \cos{(2{\theta}_2)}]+u_0 \delta({\theta}_1-{\theta}_2),
\label{ham1}
\end{equation}
where $I=m{\rho}^2$, $L_i$ and ${\theta}_i$ are the angular momentum and angle, respectively,  of the $i$th particle ($i=1,~2$), $\kappa_0$ is the lattice amplitude, and $u_0$ is the strength of the point contact pseudopotential interaction between the bosons. 

It is useful to write the Hamiltonian in terms of dimensionless parameters $(L'_i, {\theta}_i')$. For particles interacting with the optical lattice in Eq. (\ref{ham1}), the angular momentum transfer  occurs in discrete units ${\Delta}L=2{\hbar}$. We therefore define $L'_i=\frac{L_i}{2\hbar }$, and ${\theta}'_i=2 {\theta}_i$. Thus, $H'_i=\frac{H}{4\hbar\omega_r}$, $\kappa'=\frac{\kappa}{4\hbar\omega_r}$, $u'_0=\frac{u_0 }{2\hbar\omega_r}$ and $t'=4\omega_r t$, where $\omega_r=\frac{\hbar }{2I}$ is the recoil frequency. We also scale all other frequencies as $\omega'=\frac{\omega}{4\omega_r}$. We  then drop the primes on the dimensionless parameters and obtain the dimensionless Hamiltonian 
\begin{equation}
H_0=L^2_1 + L^2_2+\kappa_0 [ \cos{{\theta}_1}+ \cos{{\theta}_2}] + u_0 \delta({\theta}_1-{\theta}_2).
\label{eq:hamscale}
\end{equation}
We use as a basis set, the eigenstates of the angular momentum operator 
${\hat L}|n{\rangle}=n{|n \rangle}$ or ${\langle}{\theta}|n{\rangle}=\frac{1}{\sqrt{2\pi}}{\rm e}^{in{\theta}}$ where integers $n$ range over the values $-\infty{\leq}n{\leq}\infty$ . 

\begin{figure}[hbt]
\hspace*{-0.8in}
\ \psfig{file=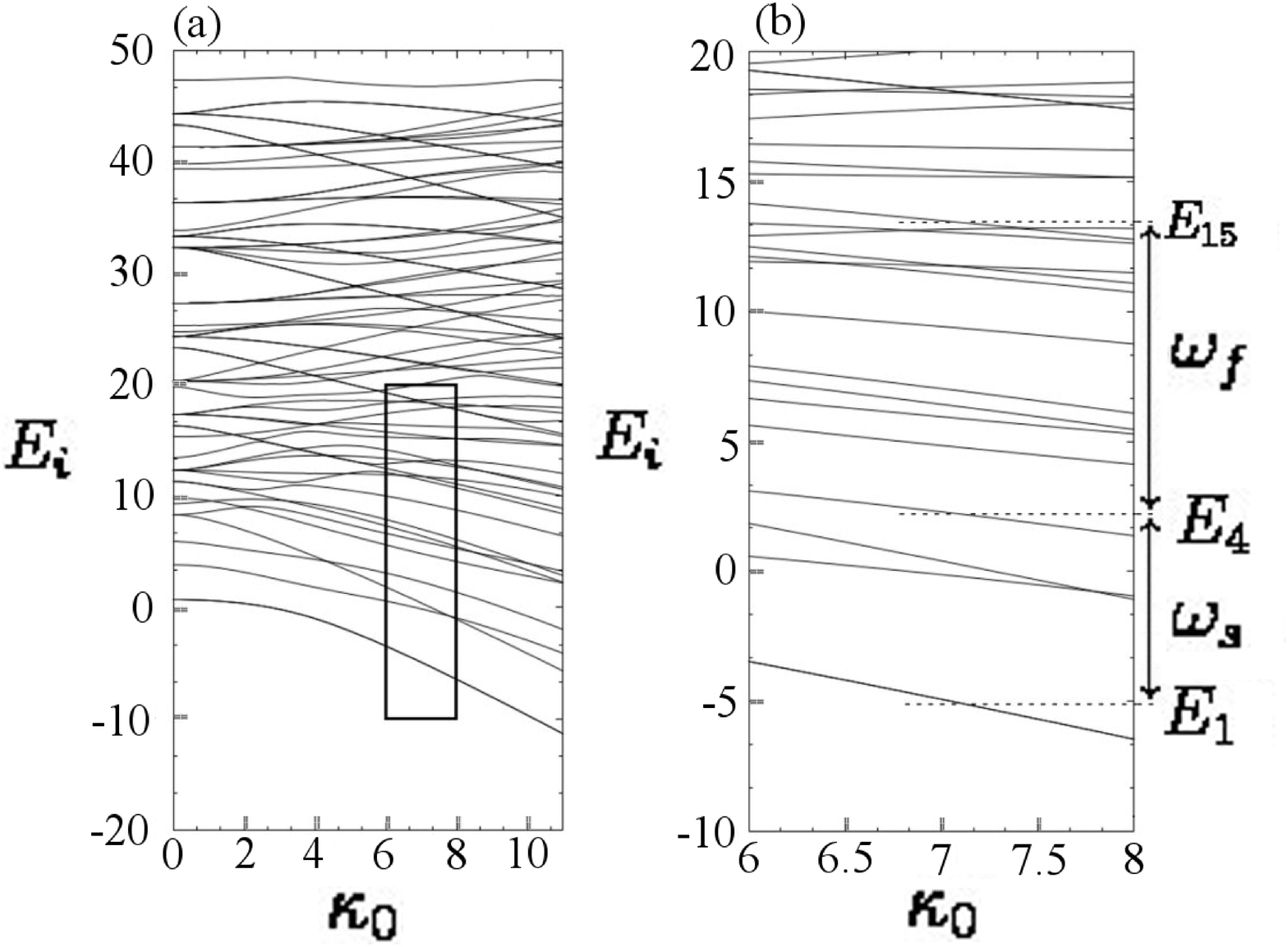,height=5.0in,width=6.0in}
\caption{ Energy curves of the first $66$ lowest-energy states of both even and odd parities for the two-boson system.  (a) The energy levels plotted as a function of  $\kappa_0$ for interaction amplitude $u_0=23.0$. The region of interest is highlighted by a box. (b) Magnified view of the region of interest boxed in Fig. 1.a.  The levels being connected by STIRAP for this particular value of $\kappa_0$ are indicated. The value of $\kappa_0$ has been adjusted so that $\frac{\omega_f}{\omega_s}=\frac{3}{2}$.}
\label{fig:energylevels}
\end{figure}

\begin{figure}[hbt]
\ \psfig{file=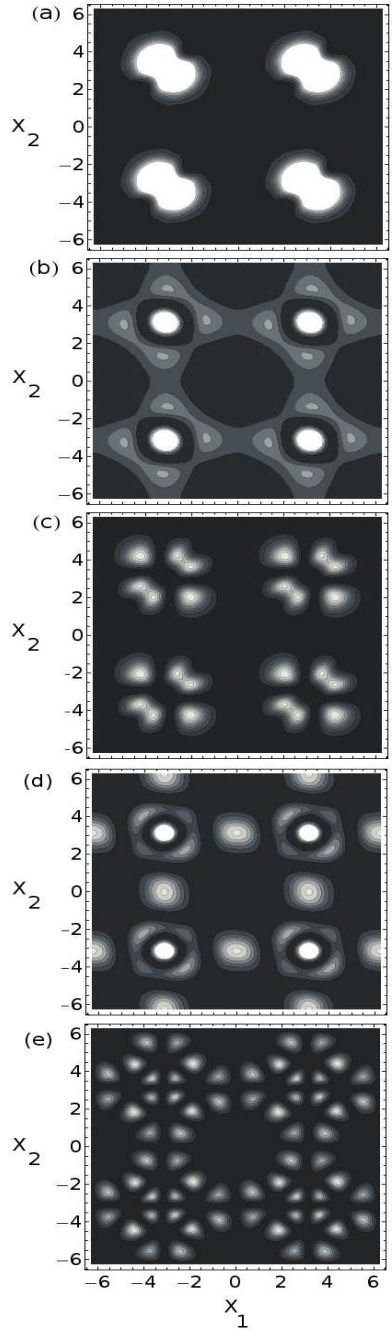,height=7.0in,width=2.0in}
\caption{Wavefunction plots for states (a) $|E_1{\rangle}$, (b) $|E_3{\rangle}$, (c) $|E_4{\rangle}$, (d) $|E_5{\rangle}$, (e) $|E_{15}{\rangle}$.}
\label{fig:wavefunctions }
\end{figure}

We numerically diagonalize the Hamiltonian in Eq. (\ref{eq:hamscale}) using a nonadaptive finite element method. The 2-particle boson states are obtained by taking  symmetrized products  of single particle states:
\begin{equation}
{\langle}{\theta}_1,{\theta}_2\vert n_1,n_2{\rangle} ^{(s)}=\frac{1}{\sqrt{2}} 
[{\langle}{\theta}_1|n_1\rangle{\langle}{\theta}_2|n_2\rangle +{\langle}{\theta}_1|n_2\rangle{\langle}{\theta}_2|n_1\rangle ].
\label{eq:symm }
\end{equation}
These states are then used to create a Hamiltonian matrix from Eq.~\eqref{eq:hamscale}. The eigenvalues $E_{\alpha}$ and eigenvectors $\vert E_{\alpha}\rangle$ of the Hamiltonian matrix were determined numerically using the appropriate subroutine for diagonalizing real symmetric matrices in the GNU Scientific Library~\cite{galassi:gsl}. Figure \ref{fig:energylevels}.a shows the energy levels $E_1$ through $E_{66}$. Figure \ref{fig:energylevels}.b gives of magnified view of these levels. 
Figures~\ref{fig:wavefunctions }.a through~\ref{fig:wavefunctions }.e are the probability distribution plots for the states $\langle {\theta}_1, {\theta}_2 | E_1 \rangle$, $\langle {\theta}_1, {\theta}_2 | E_3 \rangle$, $\langle {\theta}_1, {\theta}_2 | E_4 \rangle$, $\langle {\theta}_1, {\theta}_2 | E_5 \rangle$, and $\langle {\theta}_1, {\theta}_2 | E_{15} \rangle$ respectively at the value  $\kappa_0=7.287781$, the lattice amplitude we use in subsequent sections. We chose these states because they have the largest coupling ${\langle}E_{\alpha'}|\cos({\theta}_i)|E_{\alpha}{\rangle}$ and lead to robust STIRAP transitions. 

\section{Induced Transitions in the Interacting System}
\label{sec:qmechint}
Our goal will be to transition the bosons from the ground state $|E_1{\rangle}$ of the optical lattice to the state $|E_{15}{\rangle}$, using $|E_4{\rangle}$ as the intermediate state. We will accomplish this  by perturbing the system with two Gaussian shaped radiation pulses, the first pulse with carrier frequency ${\omega}_f=E_{15}-E_4$ and the second pulse with carrier frequency ${\omega}_s=E_4-E_1$.  In the presence of these pulses, the Hamiltonian takes the form
\begin{equation}
H(t)=L^2_1 + L^2_2+\kappa(t) [ \cos{{\theta}_1}+ \cos{{\theta}_2}] + u_0 \delta({\theta}_1-{\theta}_2),
\end{equation}
where
\begin{equation}
\kappa(t) ={\kappa}_0+ {\lambda}_f(t) \cos({\omega}_ft)+{\lambda}_s(t) \cos({\omega}_st).
\end{equation}
Here, the Gaussian amplitudes $ {\lambda}_f(t) ={\lambda}_0{\rm exp}[-(t-t_f^o)^2/4t_d^2]$ and $ {\lambda}_s(t) ={\lambda}_0{\rm exp}[-(t-t_s^o)^2/4t_d^2]$.
 
As shown in references  \cite{na-reichl:mol-rot}, \cite{holder:reichl:2res},  \cite{roy:reichl:dblwell} and \cite{na-reichl:pbox},  it is possible to use Floquet theory to analyze the effect of the radiation pulses on the system as they pass through the system.  The only requirement is that the pulse envelopes 
${\lambda}_f(t)$ and $ {\lambda}_s(t)$  be slowly varying compared to the carrier wave periods 
$2\pi/{\omega}_f$ and $2\pi/{\omega}_s$,  respectively.  We can then break the time evolution into $N$ narrow time windows, the $k$th window  centered at a time $t=t_{fix}^k$. In the $k$th time window we can write the Hamiltonian in the form
\begin{equation}
H(t;t_{fix}^k)=L^2_1 + L^2_2+\kappa(t;t_{fix}^k) [ \cos{{\theta}_1}+ \cos{{\theta}_2}] + u_0 \delta({\theta}_1-{\theta}_2).
\end{equation}
The Hamiltonian $H(t;t_{fix}^k)$ for the $k$th time window is time-periodic  if ${\omega}_f$ and ${\omega}_s$ are commensurate and  we can use Floquet theory to analyze the behavior of the system in that time window.  The value $\kappa_0=7.287781$ was chosen to achieve the commensurability $\omega_f/\omega_s = 3/2$, with $\omega_f=10.896058668420753$.  The period of the Hamiltonian $H(t;t_{fix}^k)$ is then given by $T=\pi \left(\frac{3}{\omega_f}+\frac{2}{\omega_s}\right)$. 

The Floquet Hamiltonian, for the $k$th time window, is given by $H_F^k(t)=H(t;t_{fix}^k)-i\frac{\partial}{\partial t}$ (in dimensionless units).  $H_F^k(t)$ is Hermitian and has eigenvalues ${\Omega}_{\nu}$ and eigenvectors $|{\phi}_{\nu}(t){\rangle}$.  The eigenvalues are defined modulo $2\pi/T$ and the eigenvectors   are time-periodic with period $T$.  The quantum state $|{\psi}^k(t){\rangle}$, which is a solution to the Schr\"odinger equation $i\frac{\partial \vert \psi^k(t) \rangle}{\partial t}=H(t;t_{fix}^k)|{\psi}^k(t){\rangle}$ can be expanded in a spectral decomposition 

\begin{equation}
 |\psi(t)\rangle={\sum_{\nu}}A_{\nu}e^{-i\Omega_\nu t}|\phi_\nu(t)\rangle,
\end{equation}
where $A_{\nu}$ gives the contribution of the ${\nu}$th Floquet state to the evolution of the system in a given time-window.  The Floquet evolution operator, $U_F(T)$, in the basis of symmetrized 2-boson states $\vert n \rangle \equiv \vert n_1,n_2\rangle^{(s)}$ (see Eq.~\ref{eq:symm }), is given by
\begin{equation}
\langle n \vert U_F(T)\vert m \rangle = \sum_{\nu} e^{-i\Omega_{\nu}T}\langle n \vert \phi_{\nu}(0)\rangle\langle \phi_{\nu}(0)\vert m \rangle.
\end{equation}
The Floquet evolution matrix is constructed as follows. To obtain the $m$th column of  $\langle n \vert U_F(T)\vert m \rangle$,  choose initial conditions ${\langle}m'|{\psi}(0){\rangle}={\delta}_{m',m}$ and integrate the Schr\"odinger equation from time $t=0$ to time $t=T$.  The state at time $t=T$ is the $m$th column of the Floquet evolution matrix.  For our system, the integration was  done using an $8^{th}$ order Runge-Kutta Prince Dormand algorithm~\cite{rkutta:pd} from the GNU Scientific Library~\cite{galassi:gsl}, and the Floquet evolution matrix was diagonalized using a parallelized LAPACK library through the Scalable Library for Eigenvalue Problem Computations (Slepc)~\cite{slepc}. The eigenvalues, $e^{-i\Omega_{\nu}T}$ can be used to obtain $\Omega_{\nu}$.  We obtained the Floquet eigenvalues and eigenvectors for each time window.  The eigenstates in neighboring time windows will be approximately orthonormal and we can use this fact to follow each Floquet eigenstate and eigenvalue as the pulses move through the system. 

\begin{figure}[hbt]
\vspace*{-0.1in}
\ \psfig{file=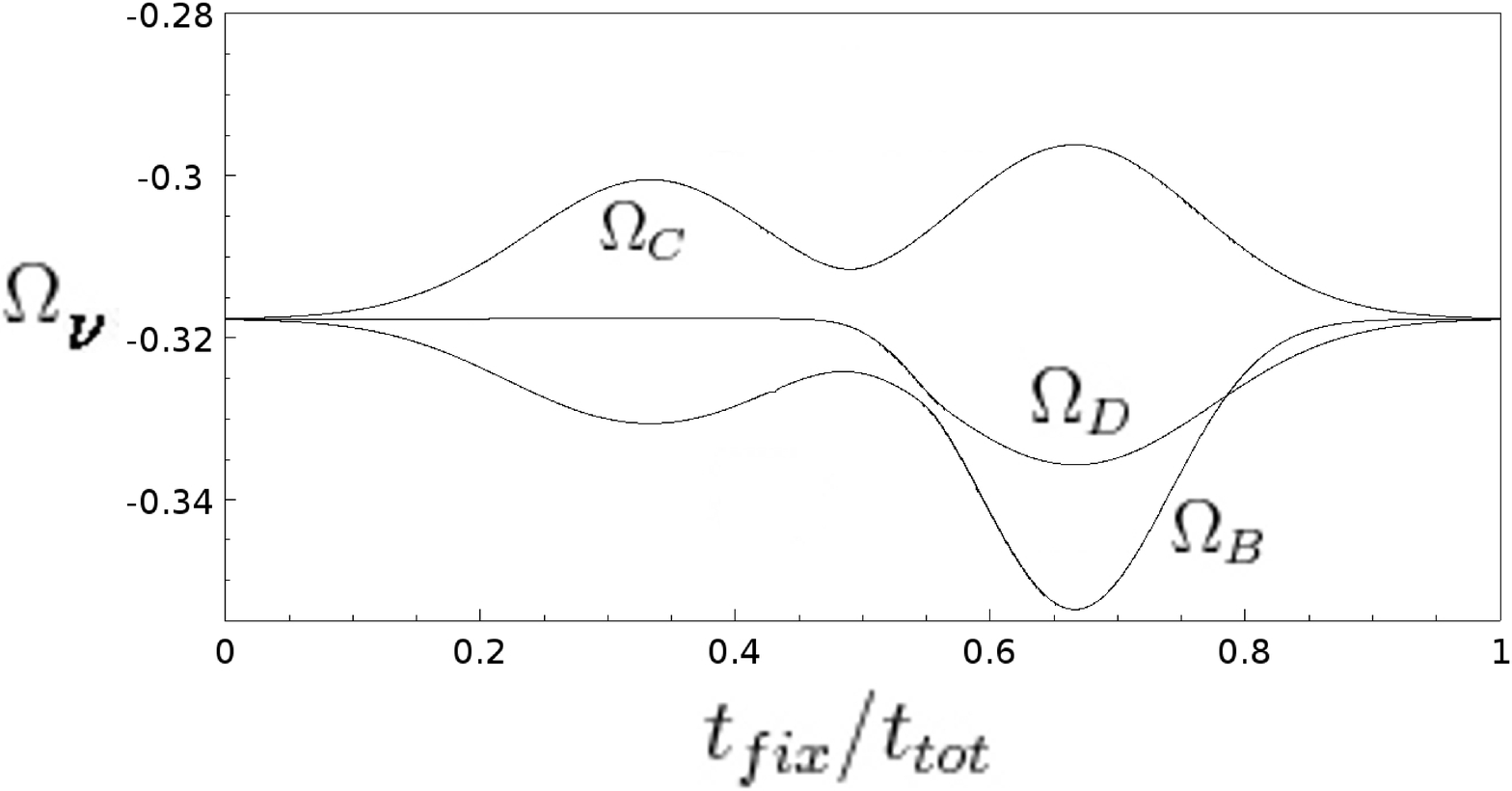,height=3.5in,width=4.6in}
\caption{Floquet quasienergies $\Omega_{\nu}$ as a function of adiabatic time $t_{fix}/t_{tot}$ for $\lambda^0=0.2$. The labels $\Omega_{B}$, $\Omega_{C}$, and $\Omega_{D}$  denote the floquet quasienergy curves corresponding to the Floquet eigenstates $|\phi_{B}\rangle$, $|\phi_{C}\rangle$, and $|\phi_{D}\rangle$,  respectively.}
\label{fig:floquet_quasi_0.2}
\end{figure}

We chose the following parameters for the Gaussian pulses;  $\lambda_0=0.2$, $t_f=(1/3) t_{tot}$, $t_s=(2/3)t_{tot}$, and $t_d=(1/14)t_{tot}$. Here $t_{tot}$ defines the total time scale for both pulses, and $t_{fix}$ is expressed in units of $t_{tot}$ unless otherwise stated. Figure~\ref{fig:floquet_quasi_0.2} shows the Floquet eigenvalues of the relevant Floquet eigenstates as the system evolves in adiabatic time $t_{fix}$. The relevant eigenstates are the ones isomorphic to the states connected by the STIRAP pulses viz. $|E_1\rangle$, $|E_4\rangle$, and $|E_{15}\rangle$. We notice that the Floquet eigenvalues are degenerate at $t_{fix}=0$ and $t_{fix}=t_{tot}$ as expected.  The Floquet states and corresponding eigenvalues are labeled alphabetically as follows:
\begin{enumerate}
\item
The eigenphase whose corresponding Floquet eigenstate is supported by the undriven state $\frac{1}{\sqrt{2}}\left[|E_4\rangle - |E_{15}\rangle \right]$ at $t_{fix}=0$ is labeled as $\Omega_B$ and the Floquet eigenstate as $|\phi_B\rangle$.
\item
The eigenphase whose corresponding Floquet eigenstate is supported by the undriven state $\frac{1}{\sqrt{2}}\left[|E_4\rangle + |E_{15}\rangle \right]$ at $t_{fix}=0$  is labeled as $\Omega_C$ and the Floquet eigenstate as $|\phi_C\rangle$.
\item 
The eigenphase whose corresponding Floquet eigenstate is supported by the undriven ground state $|E_1{\rangle}$  at $t_{fix}=0$ is labeled as $\Omega_D$ and the Floquet eigenstate as $|\phi_D\rangle$.
\end{enumerate}

\begin{figure}[hbt]
\ \psfig{file=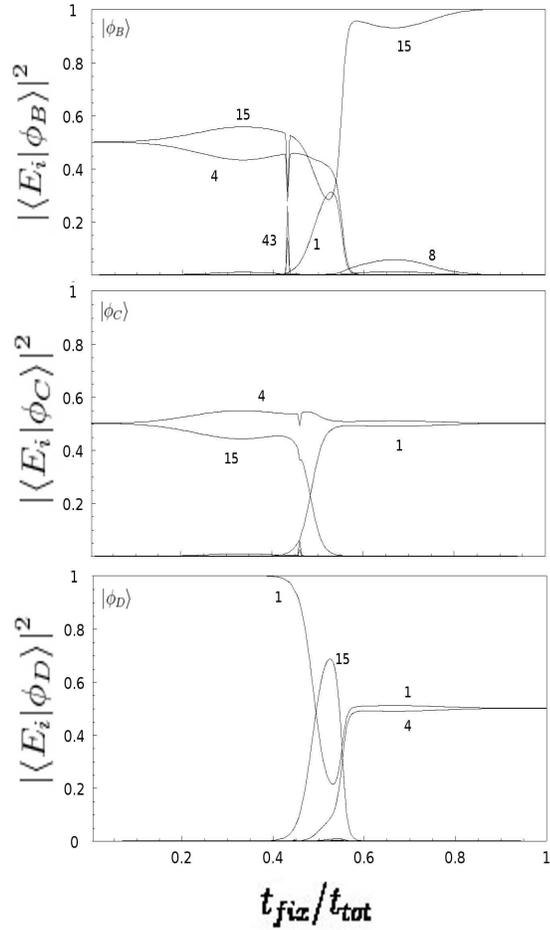,height=5.0in,width=3.0in}
\caption{Plot of the Floquet eigenfunctions  in the undriven Hamiltonian representation ($\langle E_i|\phi_{B-D}\rangle$). The components of the Floquet states in each energy level are numbered. Note the influence of the avoided crossings in $|\phi_D\rangle$.}
\label{fig:floquet_states_0.2 }
\end{figure}

The evolution of the Floquet eigenvalues $\Omega_B$, $\Omega_C$, and $\Omega_D$ are shown in Fig. \ref{fig:floquet_quasi_0.2} as the pulses pass through the system.      If  the system  evolves adiabatically, it stays in the state $|\phi_D\rangle$, but the amount of support it receives from each of the states $|E_j{\rangle}$ will change at each avoided crossing that $|\phi_D\rangle$ encounters. In Fig. \ref{fig:floquet_quasi_0.2}, there appears to be a  traditional $3$-level avoided crossing  at  about $t_{fix}\simeq 0.5$ $t_{tot}$. However, there are  additional avoided crossings involving  $\Omega_D$ that can  change $|\phi_D\rangle$.   In Fig. \ref{fig:floquet_states_0.2 }, we show the dependence of the three Floquet states  $|\phi_B\rangle$, $|\phi_C\rangle$, and $|\phi_D\rangle$ on the energy eigenstates 
$|E_j{\rangle}$ as the pulses pass through the system. We see that $|\phi_D\rangle$ starts out fully supported by $|E_1{\rangle}$ then partially changes its support from $\vert E_1\rangle$ to $\vert E_{15}\rangle$ and finally, after the pulses have passed, ends with equal support from $|E_1{\rangle}$ and $|E_4{\rangle}$.   State $|\phi_B\rangle$, on the other hand, begins with equal support from $|E_1{\rangle}$ and $|E_4{\rangle}$ and ends totally supported by $|E_{15}{\rangle}$.  A truly adiabatic evolution of the pulses, which would keep the system in Floquet state $|\phi_D\rangle$ the whole time, would not accomplish our goal of transitioning the system from $\vert E_1{\rangle}$ to $\vert E_{15}{\rangle}$.  The problem arises because additional avoided crossings occur that pull the system off the traditional STIRAP path.  
 The influence of an additional small  avoided crossing between $\Omega_B$ and $\Omega_D$ at about $t_{fix}=0.55$ is clearly  seen.  This additional small avoided crossing is a manifestation of classical chaos in the quantum dynamics similar to the ones seen for the double well system in~\cite{roy:reichl:dblwell}.
  
The dynamics can be analyzed in more detail by using the Landau-Zener formula to calculate the probability of a transition at an avoided crossing. The probability $P_{\nu,\nu'}$ for an avoided crossing between two Floquet eigenphases $\Omega_{\nu}$ and $\Omega_{\nu'}$ to be crossed  is given by~\cite{zener:lzformula},~\cite{wittig:lzformula}
\begin{equation}
P_{\nu,\nu'}=\exp\left[-\frac{\pi ({\delta \Omega_{\nu,\nu'}})^2}{2\Gamma_{\nu,\nu'}}\right],
\label{eq:landauzener }
\end{equation}
where $\delta\Omega_{\nu,\nu'}$ is the (minimum) spacing  between $\Omega_\nu$ and $\Omega_\nu'$ at the avoided crossing and $\Gamma_{\nu,\nu'}$   is the magnitude of the \textit{diabatic} rate of change (slope) of the Floquet eigenphases. Thus,
\begin{equation}
\Gamma_{\nu,\nu'}=\frac{\bar {\Gamma}_{\nu,\nu'}}{t_{tot}} = {\biggl|} \frac{d\Omega_{\nu}}{dt} - \frac{d\Omega_{\nu'}}{dt}{\biggr|},
\label{eq:gamma }
\end{equation}
where  $\frac{d\Omega_\nu}{dt}$ is the slope of the  eigenphase curve $\Omega_\nu$ in the neighborhood of the avoided crossing. Equation~(\ref{eq:landauzener }) can be simplified to
$P_{\nu,\nu'}=\exp\left[-t_{tot}{\gamma}_{\nu,\nu'} \right]$, where ${\gamma}_{\nu,\nu'}=\frac{\pi ({\delta \Omega_{\nu,\nu'}})^2}{2{\bar \Gamma}_{\nu,\nu'}}$. 

In order for a crossing to be traversed adiabatically, $P_{\nu,\nu'}\approx 0$, and the actual time scale of the STIRAP must be adjusted accordingly. Thus, the transfer probability $P_{\nu,\nu'}$ will be very small if $t_{tot}>1/{\gamma}_{\nu,\nu'}$. 
 For the $BD$ avoided crossing, we estimate the gap $\delta \Omega_{BD}$ to be $1.984 \times 10^{-3}$, and  $\Gamma_{BD}$ to be $ 0.135$, concluding that $t_{tot}>2.184 \times 10^4$. For $^{85}Rb$ atoms confined in a one-dimensional optical lattice by detuning away from the $D_2$ transition line, the recoil frequency, $ \omega_r$ is about $24$ $KHz$~\cite{steck}. The characteristic time scale here is $1/(4\omega_r)$, or $1.03 \times 10^{-5}$ seconds. We can plug this value to the minimum value(s) of $t_{tot}$ to get the actual time. Thus, for the $^{85}Rb$ atom, we get $t_{tot}>0.225$ $sec$ for the $BD$ crossing. If the time scale for the stirap is faster than $0.225$ seconds, then the crossing is traversed diabatically and traditional 3-level STIRAP will be seen. For $^{87}Rb$ atoms in the $F=2$, $m_F=2$ state confined in the toroidal magnetic trap of Fernholtz et al, the resonance condition, as well as one dimensional confinement, are met at a radius $\sim \mu m$~\cite{fernholtz}, and can be adjusted to replicate the time scales of the optical lattice.

\begin{figure}[hbt]
\ \psfig{file=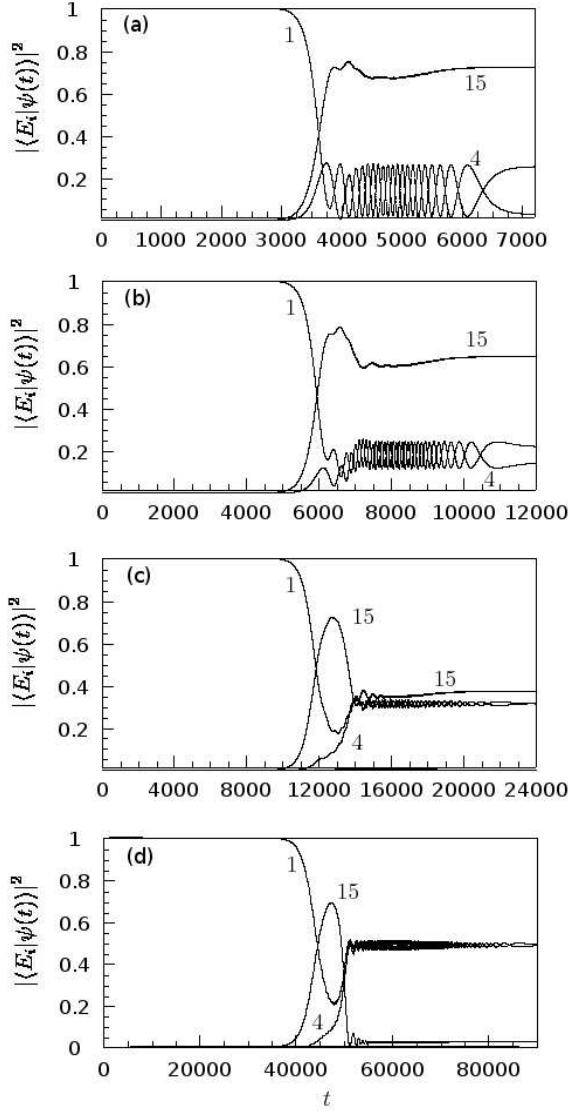,height=6.0in,width=3.0in}
\caption{Plots of $|{\langle}E_j|{\psi}(t){\rangle|}|^2$ as a function of time of the system as it evolves under the influence of STIRAP pulses from the ground state $|{\psi}(0){\rangle|}=\vert E_1\rangle$ for different values of  $t_{tot}$. All units are dimensionless. (a) $t_{tot}=7200$. (b) $t_{tot}=12,000$. (c) $t_{tot}=24,000$. (d) $t_{tot}=90,000$.}
\label{fig:timeev_groundstate }
\end{figure}

We now compare the results obtained from Floquet theory to numerical simulations of the actual dynamics of the system as it evolves through time. The full multilevel Schr\"odinger equation for this system, starting from the undriven ground state $|{\psi}(0){\rangle}=\vert E_1\rangle$, can be solved numerically for the system as it evolves from $t=0$ to $t=t_{tot}$. The rate at which the avoided crossings are traversed can be controlled by controlling $t_{tot}$. Figures~\ref{fig:timeev_groundstate }.a through~\ref{fig:timeev_groundstate }.d show the numerical time evolution of the wavefunction of the two-boson system $\vert \psi(t)\rangle$, starting from the ground state $\vert E_1\rangle$. For small values of $t_{tot}$, the system traverses the $BD$ avoided crossing diabatically, and a near-complete population transfer to $\vert E_{15}\rangle$ occurs due to the preceding 3-level avoided crossing, replicating the traditional STIRAP process. This can be seen in Figure~\ref{fig:timeev_groundstate }.a, where $t_{tot}=7200$. As we increase $t_{tot}$ towards (and beyond) $2.184 \times 10^4$, the time evolution approaches that of the Floquet state $\vert \phi_D\rangle$ as shown in Fig~\ref{fig:floquet_states_0.2 }. Figure~\ref{fig:timeev_groundstate }.d shows $\vert \psi(t)\rangle$ as it evolves in time for $t_{tot}=90,000$. The centroids of the components $\langle E_i \vert \psi(t)\rangle$ are identical to the components  $\langle E_i \vert \phi_D\rangle$ in Fig~\ref{fig:floquet_states_0.2
 }. The oscillations in the probability occur due to nonadiabatic effects as demonstrated by Berry~\cite{berry:base}, and decrease in amplitude as we move further into the adiabatic regime. 
\section{Conclusion}
\label{conclusion}
We have analyzed the dynamics of interacting two-boson systems for ultracold alkali metal atoms in an optical lattice with periodic boundary conditions.  We have demonstrated the feasibility of a controlled excitation of the system into a higher energy state using STIRAP, induced by time dependent modulations of the optical lattice. 

For sufficiently large amplitude modulations, the effects of the underlying classical dynamics \cite{reichl}  start to manifest themselves through small avoided crossings between the involved Floquet eigenphases. The STIRAP pulses were tuned to connect very high energy states (the final state being the fifteenth energy level). Avoided crossings between the other Floquet states connected  cause the outcome  to differ from  traditional three level STIRAP.  By traversing these additional small avoided  crossings  diabatically, in order to eliminate their effect on the system, we obtain the outcome expected for a three-level STIRAP process. 
\section{Acknowledgments}
The authors wish to thank the Robert A. Welch Foundation (Grant No. F-1051) for support of this work. The  authors also thank the Texas Advanced Computing Center (T.A.C.C.) at the University of Texas at Austin for the use of their high-performance distributed computing grid.

\pagebreak
\listoffigures
\pagebreak
%
%

\end{document}